# Data-Driven Calibration of Multi-Fidelity Multiscale Fracture Models via Latent Map Gaussian Process

Shiguang Deng [a], Carlos Mora [b], Diran Apelian [a], Ramin Bostanabad [b, *]

[a] ACRC, Materials Science and Engineering, University of California, Irvine, CA, USA
[b] Mechanical and Aerospace Engineering, University of California, Irvine, CA, USA

**Abstract**

Fracture modeling of metallic alloys with microscopic pores relies on multiscale damage simulations which typically ignore the manufacturing-induced spatial variabilities in porosity. This simplification is made because of the prohibitive computational expenses of explicitly modeling spatially varying microstructures in a macroscopic part. To address this challenge and open the doors for fracture-aware design of multiscale materials, we propose a data-driven framework that integrates a mechanistic reduced-order model (ROM) with a calibration scheme based on random processes. Our ROM drastically accelerates direct numerical simulations (DNS) by using a stabilized damage algorithm and systematically reducing the degrees of freedom via clustering. Since clustering affects local strain fields and hence the fracture response, we calibrate the ROM by constructing a multi-fidelity random process based on latent map Gaussian processes (LMGPs). In particular, we use LMGPs to calibrate the damage parameters of an ROM as a function of microstructure and clustering (i.e., fidelity) level such that the ROM faithfully surrogates DNS. We demonstrate the application of our framework in predicting the damage behavior of a multiscale metallic component with spatially varying porosity. Our results indicate that microstructural porosity can significantly affect the performance of macro components and hence must be considered in the design process.

## 1. Introduction

Multiscale models are increasingly employed to quantify the effects of manufacturing-induced microscopic defects on the performance of macroscopic components. In such models, a microstructure or a representative volume element (RVE) is associated with each integration point (IP) of the discretized macrostructure. Traditional multiscale simulations use the finite element method (FEM) to solve the nonlinear equilibrium equations at both scales where macroscopic deformation gradients $\mathbf{F}^M$ and RVE effective stress $\boldsymbol{\sigma}^M_{FEM}$ are exchanged between the two scales at each iteration, see Figure 1(a). A major challenge associated with such nested simulations is the computational expenses which prohibitively increase in the presence of nonlinear microscale deformations that involve damage. Reducing these costs holds the key to understanding the relation between microscopic defects and components' fracture behavior and, in turn, guiding the "design for fracture" process. To this end, we propose a data-driven framework that has two major components: (1) a mechanistic reduced-order model (ROM) with an adjustable degree of fidelity, and (2) a multi-fidelity modeling and calibration scheme based on latent map Gaussian processes (LMGPs) [1, 2]. Integration of these two components enables us to build calibrated multi-fidelity ROMs that can simulate the damage behavior of multiscale materials with spatially varying microstructures.

The rest of our paper is organized as follows. In Section 2, we review existing works on reduced-order modeling and discuss the research gaps that we aim to address. The overview and technical details of our approach are provided in Sections 3 and 4, respectively. We evaluate the performance of our approach in Section 5 and conclude the paper in Section 6.

## 2. Background on reduced-order modeling

Mechanistic ROMs are increasingly employed to accelerate nonlinear materials modeling by using a combination of methods from linear algebra and machine learning that result in reducing the number of unknown variables that characterize, e.g., microstructural strain and stress fields. Transformation field analysis (TFA) and its successor nonuniform transformation field analysis (NTFA) are two of the earliest ROMs [3–5]. These two methods approximate plastic strain as either piecewise constants or spatially varying orthonormal eigenstrains which are pre-selected in an offline stage. These eigenstrains evolve in the online stage based on pre-defined analytical functions that involve thermodynamic forces and potentials.

Clustering-based ROMs are recent techniques that decompose microstructure domains into a set of clusters whose interactions and deformations are modeled. For instance, the self-consistent analysis (SCA) [6] lumps material points with similar elastic responses and then quantifies cluster-to-cluster interactions by the incremental Lippmann-Schwinger equation. Finite element-based cluster analysis [7] approximates the microstructural effective responses by following the cluster minimum complementary energy principle. Deflated clustering analysis (DCA) [8] agglomerates close-by material integration points (IPs) into clusters and the cluster-wise quantities of interests are computed in a multi-grid fashion where unknown variables are projected back and forth between different meshes. In this work, we use cluster-based ROMs as they provide higher efficiency and versatility compared to other methods such as TFA.

---

*Corresponding author.
 Email address: raminb@uci.edu.

Successful application of any ROM depends on two primary factors: (i) the coarsening degree (e.g., the chosen number of clusters) which makes a tradeoff between *fidelity* level and computational costs, and (ii) the calibrated material properties. Both of these factors depend on the microstructure as well as the properties of interests. For example, accurate prediction of the damage behavior requires different damage parameters and a number of clusters for the two microstructures in Figure 1(a). In particular, given a desired level of accuracy with respect to high-fidelity direct numerical simulations (DNS), the analysis of the more complex microstructure in Figure 1(b) generally requires more clusters (i.e., less coarsening or data reduction).

Regarding the second requirement of the successful application of ROMs, we note that accurate prediction of damage behavior necessitates the calibration of material properties to account for the diffusive stress and strain fields of any ROM. The diffusion typically depends on the microstructure topology, and it unrealistically increases the tolerance of the material system to localized phenomena. The superficial increase of material strength upon clustering, therefore, must be counteracted in ROM to ensure solution accuracy. We clarify that, in this paper, we use the word 'diffusion' to exclusively refer to the artificially strengthening of material clusters in ROMs (which aim to capture the homogenized behavior of the material encompassed in the cluster) and not the transfer of matter by diffusion.

While calibrating material properties plays a vital role in ensuring that ROMs can be reliably used in multiscale simulations, there is still a lack of systematic approaches that dispense with manual calibration which is time-consuming and suboptimal. As explained in the next sections, our contribution is to develop a data-driven framework to automate the calibration process of ROMs with any fidelity level as a function of material morphology.

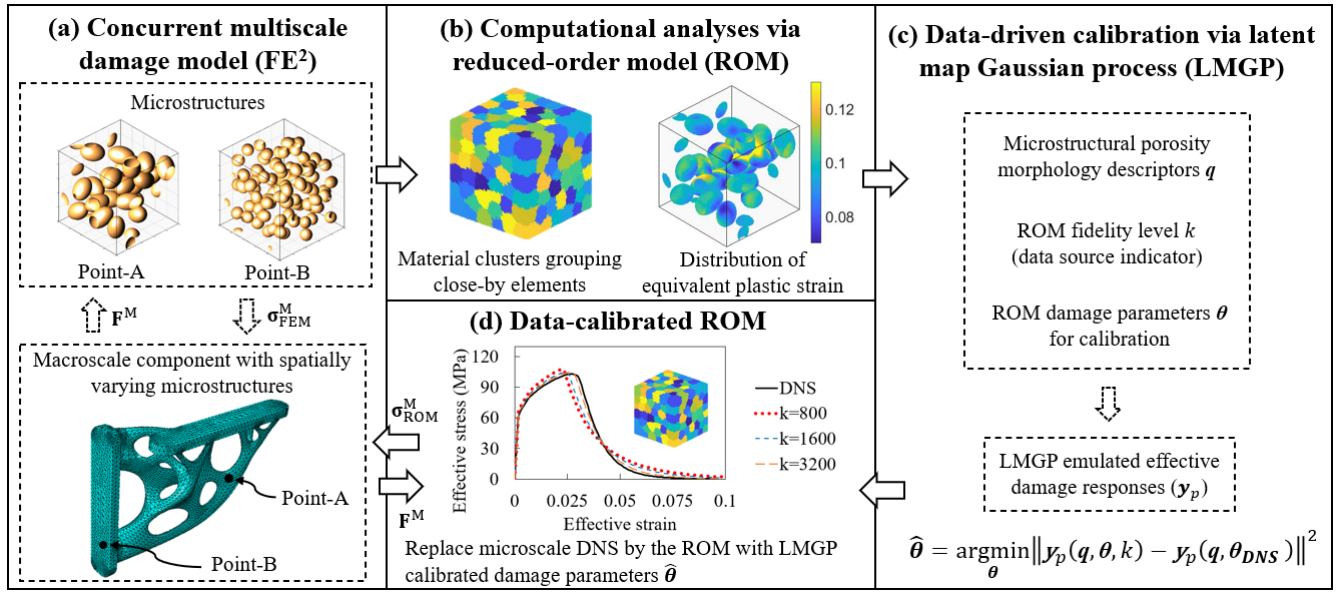

**Figure 1 Proposed data-driven framework for multiscale damage modeling:** LMGP creates a multi-fidelity emulator for the ROMs and DNS. It is then used in an inverse optimization to determine the damage parameters that must be used in ROMs such that they approximate DNS as closely as possible conditioned on the microstructure. Upon this calibration, a multiscale simulation is run where ROMs are used at the microscale.

## 3. Overview of the proposed framework

Our framework relies on two primary components for damage modeling in multiscale metals with porosity: a novel cluster-based ROM and LMGP-based calibration which are detailed in Sections 4.3 and 4.4, respectively. We provide an overall description of these components in this section.

The ROM surrogates DNS and estimates the stress field in a microstructure under arbitrary displacement boundary conditions that may result in plasticity and damage. The fidelity of the ROM is determined by the user-defined parameter $k$ which indicates the number of clusters and balances costs and accuracy.

As argued in Section 2, the material properties that must be used in ROM should be different than the true values that are used in DNS, i.e., the ROM requires calibration. This difference depends on both the microstructure complexity and, more importantly, $k$. Hence, we use a data-driven approach that relies on emulation via an LMGP to calibrate the material properties for ROMs. In particular, we use the trained LMGP emulates ROM and DNS to answer the following question:

*Given $k$ and one microstructure, what damage parameters should be used in the ROM such that it predicts the same fracture response as DNS which uses the true damage parameters?*

We answer the above question by solving an inverse optimization problem whose objective function relies on LMGP, see Figure 1(c). To make the optimization problem tractable, we make two mild assumptions. Firstly, we consider a small set of integer values for $k$, i.e., we assume $k = 800, 1600,$ or $3200$ but more values can be used within our framework. As shown in Section 4.3, all these values are much smaller than the number of elements in a typical mesh used in DNS and hence result in massive data reduction or coarsening. Secondly, the very high dimensional morphology of microstructures can be represented with a reduced set of quantitative descriptors that in our case characterize the geometry and spatial distribution of the pores.

We note that the clustering-based ROMs are new methods developed in recent years. Even though ROMs dramatically improve simulation efficiency, the number of clusters in ROMs is currently chosen in an ad-hoc manner and there still lacks theoretical proof on the criteria to choose the number of clusters for arbitrary material systems. This is because material systems can be very different in local morphology, material composition/property, defect types/distributions, etc. More complex material systems generally require higher fidelity models and therefore more clusters. The goal of our work is to replace this manual approach for selecting the number of clusters with an automated method. In this work, we start with 800 clusters and then doubled this number twice to get 1600 and 3200 clusters. We intentionally did *not* study lower/higher clusters because we aim to develop a rather general calibration scheme that is not sensitive to the chosen cluster numbers.

To build the LMGP, we generate the training samples by the design of experiments (DoE) where the inputs are microstructural descriptors and calibration parameters that control the damage behavior. For sample $i$, we first use a reconstruction algorithm to build the microstructure corresponding to the $i^{th}$ set of descriptors. Then, we calculate the fracture response of the $i^{th}$ microstructure via a simulator (i.e., DNS or one of the ROMs) while using the $i^{th}$ set of damage parameters. When obtaining the responses, we select the simulator based on its computational costs, i.e., the frequency of using a simulator is inversely proportional to its costs (e.g., we employ an ROM with small $k$ much more than DNS or an ROM with large $k$).

It is noted that the optimization problem uses LMGP rather than a traditional Gaussian process (GP) since we view the data source indicator as a categorical input rather than a quantitative one, see Figure 1(c). This choice is justified since alternating the data source (e.g., DNS vs. ROM with $k = 800$ vs. ROM with $k = 3200$) encodes the diffusive nature of strain-stress fields which cannot be readily characterized with quantitative inputs. Hence, our treatment of data source motivates the use of LMGP and greatly simplifies the emulation as it eliminates the manual conversion of the source label to a quantitative variable.

Once LMGP is built, we are ready to run a multiscale simulation where ROMs are used at the microscale instead of DNS, see Figure 1(d). We first assign spatially varying microstructures to the IPs on the macro-component. Then, based on the complexity of the microstructures and any prior knowledge (if available) on the macro-locations where excessive deformations can occur (e.g., near sharp corners), we choose the $k$ values for ROM. Next, we use the trained LMGP to assign the damage parameters that must be used at $i^{th}$ macro-IP given the $k$ and microstructure assigned to it. Upon this assignment, we conduct the multiscale simulation to find the performance of the macro-component while considering microstructural porosity.

## 4. Technical details

We first provide the details on damage modeling with our ROM in Sections 4.1 through 4.3. Then, we elaborate on the training process of LMGPs in Section 4.4 and explain our optimization-based calibration algorithm in Section 4.5.

### 4.1. Stabilized micro-damage model

Damage includes strain-softening which causes convergence issues in implicit time integration schemes. To address this issue, we use a stabilized damage model [9] to simulate microstructural effective responses during fracture progression. This model decouples damage evolution from elasto-plasticity by introducing three reference RVEs that share state variables with the original damaged RVE. By tracing the elasto-plasticity in one of the referenced RVEs via a classic implicit scheme, the effective fracture stress and states can be mapped to the damaged RVE.

The homogenized damage stress in an arbitrary RVE can be written as:

$$\mathbf{S}_M^d = \mathbb{C}_M^d : \mathbf{E}_M^{el} = \mathbb{C}_M^d : (\mathbf{E}_M - \mathbf{E}_M^{pl}) \qquad (1)$$

where $\mathbf{S}_M^d$ represents the effective damage stress, $\mathbb{C}_M^d$ is the homogenized damaged tangent modulus matrix, $\mathbf{E}_M$, $\mathbf{E}_M^{el}$ and $\mathbf{E}_M^{pl}$ are the RVE effective strain, elastic strain, and plastic strain, respectively. The subscript M indicates that the variable is a macroscopic quantity. The symbol ':' represents the double dot product that contracts a pair of repeated indices.

The first reference RVE shares the same elasto-plastic deformation as the original RVE but is not damaged. Its effective stress is therefore computed as:

$$\mathbf{S}_M^1 = \mathbb{C}_M^{el} : \mathbf{E}_M^{el} = \mathbb{C}_M^{el} : (\mathbf{E}_M - \mathbf{E}_M^{pl}) \qquad (2)$$

where $\mathbf{S}_M^1$ and $\mathbb{C}_M^{el}$ represent the homogenized stress and the undamaged elastic modulus, respectively (superscript 1 refers to the first referenced RVE). By combining Equations (1) and (2), we can express the referenced stress as:

$$\mathbf{S}_M^1 = \mathbb{C}_M^{el} : (\mathbb{C}_M^d)^{-1} : \mathbf{S}_M^d \qquad (3)$$

The second reference RVE has the same effective stress ($\mathbf{S}_M^2 = \mathbf{S}_M^1$) and material property as the first RVE but is assumed to deform elastically. Thus, its effective elastic strain ($\mathbf{E}_M^{el}$) is:

$$\mathbf{E}_M^{el} = \left(\mathbb{C}_M^{el}\right)^{-1} : \mathbf{S}_M^1 = \left(\mathbb{C}_M^{el}\right)^{-1} : \mathbf{S}_M^2 \qquad (4)$$

The effective stress and strain of the second reference RVE are equivalently expressed as the volume average of its microscale stress and strain as:

$$\mathbf{S}_M^2 = \frac{1}{|\Omega|} \int_\Omega \mathbf{S}_m^2 d\Omega \qquad (5)$$

$$\mathbf{E}_M^{el} = \frac{1}{|\Omega|} \int_\Omega \mathbf{E}_{m2}^{el} d\Omega \qquad (6)$$

where $|\Omega|$ is the RVE volume, the subscript m indicates that the variable is a microscopic quantity, and the microscale stress $\mathbf{S}_m^2$ is proportional to the microscale elastic strain $\mathbf{E}_{m2}^{el}$ at any microscopic point by the elastic modulus $\mathbb{C}^{el}$ via:

$$\mathbf{S}_m^2 = \mathbb{C}^{el} : \mathbf{E}_{m2}^{el} \qquad (7)$$

The third reference RVE has the same elastic strain as the second one ($\mathbf{E}_{m3}^{el} = \mathbf{E}_{m2}^{el}$) but its modulus is assumed to be identical to the original fractured RVE as:

$$\mathbf{S}_m^3 = \mathbb{C}_m^d : \mathbf{E}_{m3}^{el} \qquad (8)$$

$$\mathbb{C}_m^d = (1 - D_m)\mathbb{C}^{el} \qquad (9)$$

where $\mathbb{C}_m^d$ is the microscale damaged tangent modulus and $D_m$ is the damage parameter at a microscopic IP. The value of $D_m$ is determined by the plastic strain states in the first reference RVE:

$$D_m(\bar{E}^{pl}; \alpha, \bar{E}^{cr}) = 1 - \frac{\bar{E}^{cr}}{\bar{E}_{m1}^{pl}} \exp(-\alpha(\bar{E}_{m1}^{pl} - \bar{E}^{cr})) \qquad (10)$$

where $\bar{E}^{pl}$ is the equivalent plastic strain, $\bar{E}_{m1}^{pl}$ is the equivalent plastic strain at a microscale IP in the first referenced RVE, and $\bar{E}^{cr}$ is the critical plastic strain. $\alpha$ is the damage evolutionary rate parameter and a larger value of $\alpha$ results in faster material degradations and rapid effective stress drop amid softening. We note that local damage is initiated ($D_m = 0$) when effective plastic strain equals the critical strain ($\bar{E}_{m1}^{pl} = \bar{E}^{cr}$) and damage reaches total rupture ($D_m = 1$) when the effective plastic strain is much larger than the critical plastic strain.

The effective damaged stress of the original RVE is assumed to be equal to the homogenized stress of the third reference RVE and is calculated as:

$$\mathbf{S}_M^d = \mathbf{S}_M^3 = \frac{1}{|\Omega|} \int_\Omega \mathbf{S}_m^3 d\Omega \qquad (11)$$

For the multiscale damage analysis in Section 5.4, the macroscale damage parameter is computed as the ratio of the norms of effective stress tensors of the original and the first reference RVE as:

$$D_M = 1 - \frac{\left\| \mathbf{S}_M^d : \mathbf{S}_M^1 \right\|}{\left\| \mathbf{S}_M^1 : \mathbf{S}_M^1 \right\|} \qquad (12)$$

where $D_M$ is the homogenized damage parameter representing the fracture status of a macroscale IP (and its associated RVE) on a macroscale component.

### 4.2. Condensation method

When using the stabilized micro-damage model of Section 4.1 in a multiscale simulation, the effective elastic tangent moduli $\mathbb{C}_M^{el}$ is needed at each macroscopic IP, see Equation (2). Since we assign spatially varying RVEs with complex morphologies to macro IPs, $\mathbb{C}_M^{el}$ needs to be computed via variational principles for each RVE [10]. This numerical procedure is needed since the constitutive laws of the RVEs are not available in closed form.

As variational calculations are expensive, we employ the condensation method [11] to compute the effective tangent modulus of an RVE. The condensation method starts by partitioning the microstructural system of equations as:

$$\begin{bmatrix} \mathbf{K}_{pp} & \mathbf{K}_{pf} \\ \mathbf{K}_{fp} & \mathbf{K}_{ff} \end{bmatrix} \begin{bmatrix} \delta \mathbf{u}_p \\ \delta \mathbf{u}_f \end{bmatrix} = \begin{bmatrix} \delta \mathbf{f}_p \\ \mathbf{0} \end{bmatrix} \qquad (13)$$

where $\delta \mathbf{u}_p$ and $\delta \mathbf{u}_f$ represent the displacement variations at the prescribed and free nodes in an RVE where the indices **p** and **f** represent the prescribed and free degrees of freedom, and $\delta \mathbf{f}_p$ is the external force on the nodes with prescribed forces. $\mathbf{K}_{pp}, \mathbf{K}_{pf}, \mathbf{K}_{fp}$ and $\mathbf{K}_{ff}$ are the corresponding partitions of the RVE's stiffness matrix.

Eliminating $\delta \mathbf{u}_f$ from Equation (13) leads to a reduced system, with a reduced stiffness $\mathbf{K}_r$ which directly relates the variations of the prescribed displacements with nodal forces:

$$\mathbf{K}_r \delta \mathbf{u}_p = \delta \mathbf{f}_p \qquad (14)$$

$$\mathbf{K}_r = \mathbf{K}_{pp} - \mathbf{K}_{pf}(\mathbf{K}_{ff})^{-1} \mathbf{K}_{fp} \qquad (15)$$

To transform $\mathbf{K}_r$ to the tangent modulus that relates variations of stress and strain, we substitute Equation (14) into the variational form of the macroscopic stress:

$$\mathbf{S}_M(\mathbf{X}) = \frac{1}{|\Omega_{0m}|} \int_{\Gamma_{0m}} \bar{\mathbf{t}}_m \otimes (\mathbf{x} - \mathbf{x}_0) d\Gamma \qquad (16)$$

where $\mathbf{x}$ and $\mathbf{x}_0$ are the microscale IPs at the deformed and original configurations, $\mathbf{S}_M$ is the macroscale stress at the macroscopic IP $\mathbf{X}$, $\bar{\mathbf{t}}_m$ is the microscale surface traction, $\Gamma_{0m}$ is the RVE boundary, and $\otimes$ denotes the tensor product between $\bar{\mathbf{t}}_m$ and the position vector ($\boldsymbol{x} - \boldsymbol{x}_0$). Upon some algebraic modifications, the homogenized tangent (elastic) modulus matrix of an RVE can be obtained as:

$$\mathbb{C}_M^{el} = \frac{1}{|\Omega_{0m}|}\left[(\mathbf{x}-\mathbf{x}_0)\otimes\mathbf{K}_r\otimes(\mathbf{x}-\mathbf{x}_0)\right]^{LT} \quad (17)$$

where 'LT' denotes the transposition between the two left indices.

We note that even though the condensation method accelerates the calculation of $\mathbb{C}_M^{el}$ for each RVE, parallel computations based on it in a multiscale analysis are memory demanding and still quite expensive. Hence, to avoid the online condensation procedure, we utilize a GP to learn the relation between microstructural morphology and effective elastic tangents for different RVEs which are pre-computed by the condensation method in an offline stage.

### 4.3. Deflated clustering analysis (DCA)

In a multiscale simulation, the elasto-plastic response of the RVEs associated with the macro IPs can be obtained via the stabilized micro-damage algorithm (see Section 4.1). These computations are very expensive and so we use the DCA method [12] to dramatically accelerate them. Compared to other clustering-based ROMs [7, 9, 13] which primarily speed up micro analyses, our method can accelerate both macro- and micro- simulations. Its high efficiency comes from the fact that (1) the degrees of freedom are significantly reduced from a large number of finite elements to a few clusters by employing material clustering techniques, and (2) the algebraic system on the reduced system has much fewer close-to-zero eigenvalues (and hence better convergence behavior) compared to the classic finite element system.

DCA uses clustering to agglomerate neighboring finite elements to a set of interactive irregularly shaped clusters. Clustering is an unsupervised machine learning technique to interpret and group similar data. Among many mature clustering algorithms [14], we adopt k-means clustering [15] in this work due to its simplicity.

We start the k-means clustering by feeding the coordinates of element centers into a feature space where cluster seeds are randomly scattered and serve as initial cluster means. Then, we assign each element to the cluster with the closest mean. Meanwhile, cluster shapes are iteratively updated to minimize the within-cluster variance, see an illustration in Figure 2.

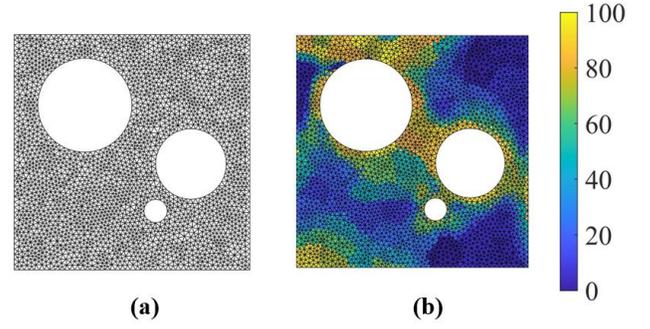

**(a)**     **(b)**

**Figure 2** Illustration of material points clustering: **(a)** a generic 2D RVE is discretized with 5000 triangle finite elements, and **(b)** the elements are grouped into 100 clusters via k-means clustering where the elements in the same cluster are indicated by the same color.

Mathematically, the clustering can be stated as the following minimization problem:

$$\mathbf{C} = \underset{\mathbf{C}}{\operatorname{argmin}} \sum_{I=1}^{k} \sum_{n\in C^I} \|\varphi_n - \bar{\varphi}_I\|^2 \quad (18)$$

where $\mathbf{C}$ represents the k-clusters with $\mathbf{C} = \{C^1, C^2, \ldots, C^k\}$. $\varphi_n$ and $\bar{\varphi}_I$ are the coordinates of the $n^{th}$ element center and the mean of the $I^{th}$ cluster, respectively.

Upon clustering, we construct a reduced mesh by connecting cluster centroids via Delaunay triangularization where topological relations are preserved by checking the connectivity between clusters. We assume the motions of cluster centroids are directly related to the grouped nodes. Specifically, the displacement of the cluster centroid $\mathbf{u}(\mathbf{x})$ is computed by interpolating the nodal displacements via the polynomial augmented radial point interpolation method [16] as:

$$\mathbf{u}(\mathbf{x}) = \sum_{i=1}^{n} R_i(\mathbf{x})a_i + \sum_{j=1}^{m} Z_j(\mathbf{x})b_j \quad (19)$$

where $a_i$ is the coefficient of the radial basis function $R_i$ at the $i^{th}$ FE node and $b_j$ is the coefficient of the polynomial basis $Z_j$. $n$ and $m$ are the number of cluster nodes and the number of polynomial basis functions, respectively. The coefficients $a_i$ and $b_j$ are determined by enforcing Equation (19) for all nodal displacements in the cluster where polynomial basis and radial coefficients are assumed to satisfy Equation (20) to ensure solution uniqueness [16]:

$$\sum_{i=1}^{n} Z_j(\mathbf{x})a_i = 0, \quad j = 1, 2, \ldots, m \quad (20)$$

We then augment the displacements of cluster centroids with rotational degrees of freedom to represent rigid body motions (three translations and three rotations in 3D) in a deflation space [19–21] where a reduced stiffness matrix is constructed with six degrees of freedom on each node. Performing nonlinear analyses on the reduced mesh and projecting the results back to the finite element nodes at the end of computations reads:

$$\mathbf{u}_i^j = \mathbf{W}_i^j \boldsymbol{\lambda}_i \tag{21}$$

where $\mathbf{u}_i^j$ is the displacement vector at the $i^{th}$ node in the $j^{th}$ cluster, $\boldsymbol{\lambda}_i$ is the rigid-body motion of the centroid of the $j^{th}$ cluster, and $\mathbf{W}_i^j$ is the deflation matrix for the $i^{th}$ node grouped in the $j^{th}$ cluster:

$$\boldsymbol{\lambda}_j = \left[ u_{jx}, u_{jy}, u_{jz}, \theta_{jx}, \theta_{jy}, \theta_{jz} \right]^T \tag{22}$$

$$\mathbf{W}_i^j = \begin{bmatrix} 1 & 0 & 0 & 0 & z_i^j & -y_i^j \\ 0 & 1 & 0 & -z_i^j & 0 & x_i^j \\ 0 & 0 & 1 & y_i^j & -x_i^j & 0 \end{bmatrix} \tag{23}$$

where $u_{jx}$ and $\theta_{jx}$ are the displacement and rotation of the $j^{th}$ cluster along $x$ axis, and $(x_i^j, y_i^j, z_i^j)$ are the relative 3D coordinates of the $i^{th}$ node with respect to the centroid of the $j^{th}$ cluster.

We note that material points are assumed to share the same stress and strain values in each cluster. Hence, the local plastic strain fields are reproduced in a diffusive manner with lower strain concentrations which, in turn, delay the onset of localized fracture. This diffusive behavior motivates the damage parameter calibration using LMGP in the next section.

### 4.4. Latent map Gaussian Process (LMGP)

GPs are widely used in many applications for emulation [20–23]. The underlying idea of GP modeling is to assume that the data originate from a multivariate normal distribution. With this assumption, GP modeling involves considering a parametric form for the mean and covariance functions of the distribution and, in turn, estimating the parameters of these functions.

Traditional GPs cannot handle categorical inputs because covariance functions rely on the (weighted) distance between inputs while categorical inputs are not typically endowed with a distance measure. To address this limitation of GPs, we have recently developed LMGPs [1] that enable GPs to handle categorical inputs such as the data source indicator in our case. As we show in Section 5.2, the learned latent space of an LMGP provides a nice diagnostic tool that can guide the analysis and design process.

Assume the observations are produced by the single-response function $\eta(s)$ which is modeled as:

$$\eta(\mathbf{s}) = f(\mathbf{s})\beta + \xi(\mathbf{s}) + \varepsilon \tag{24}$$

where $f(s) = [f_1(s), \ldots, f_h(s)]$ is a vector of predefined parametric basis functions depending on the $d_s$ dimensional input vector $s = [s_1, s_1, \ldots, s_{d_s}]^T$, $\beta = [\beta_1, \ldots, \beta_h]^T$ represent the unknown coefficients of the basis functions, $\varepsilon$ is white noise, and $\xi(s)$ is a zero-mean GP with covariance function:

$$cov(\xi(\mathbf{s}), \xi(\mathbf{s}')) = c(\mathbf{s}, \mathbf{s}') = \sigma^2 r(\mathbf{s}, \mathbf{s}') \tag{25}$$

where $c(\cdot,\cdot)$ is the covariance function, $\sigma^2$ denotes the amplitude, and $r(\cdot,\cdot)$ is the correlation function. An example $r(\cdot,\cdot)$ is the Gaussian kernel given by:

$$\begin{aligned} r(\mathbf{s}, \mathbf{s}') &= exp\left\{ -\sum_{i=1}^{d_s} 10^{w_i} (\mathbf{s}_i - \mathbf{s}_i')^2 \right\} \\ &= exp\left\{ (\mathbf{s} - \mathbf{s}')^T \boldsymbol{\Omega}_{\mathbf{s}} (\mathbf{s} - \mathbf{s}') \right\} \end{aligned} \tag{26}$$

where $\mathbf{w} = [w_1, \ldots, w_{d_s}]^T$ is the vector of roughness parameters and $\Omega_\mathbf{s} = diag(10^\mathbf{w})$. As it can be seen, $r(\cdot,\cdot)$ in Equation (26) does not accommodate categorical inputs as the distance between them is not defined.

To handle categorical inputs, LMGP maps them into a quantitative latent space which then makes it possible to use any distance-based correlation function. Specifically, let us denote the categorical inputs via $\mathbf{t} = [t_1, \ldots, t_{d_t}]^T$ where variable $t_i$ has $m_i$ different levels. Upon mapping, LMGP uses the Gaussian correlation function as:

$$r(\mathbf{u}, \mathbf{u}') = exp\left\{ -(\mathbf{s} - \mathbf{s}')^T \boldsymbol{\Omega}_{\mathbf{s}} (\mathbf{s} - \mathbf{s}') - \|\mathbf{z}(\mathbf{t}) - \mathbf{z}(\mathbf{t}')\|^2 \right\} \tag{27}$$

where $\mathbf{u} = [\mathbf{s}; \mathbf{t}]$ and $\mathbf{z}(\mathbf{t}) = [z_1(\mathbf{t}), \ldots, z_{d_z}(\mathbf{t})]^T$ is the learned $d_z$ dimensional latent variable representing a particular combination of the categorical variables. $\mathbf{z}(\mathbf{t})$ is computed by mapping the representation of each combination of the categorical variables $\boldsymbol{\tau}(\mathbf{t})$ via:

$$\mathbf{z}(\mathbf{t}) = \boldsymbol{\tau}(\mathbf{t})\mathbf{A} \tag{28}$$

where $\mathbf{A}$ is the projection matrix that is estimated during training.

Given a training dataset with $n$ samples, the LMGP parameters (i.e., $\mathbf{A}, \boldsymbol{\beta}, \mathbf{w}$, and $\sigma^2$) are estimated by maximizing the log-likelihood function:

$$\left[ \hat{\mathbf{A}}, \hat{\boldsymbol{\beta}}, \hat{\mathbf{w}}, \hat{\sigma}^2 \right] = \underset{\mathbf{A}, \boldsymbol{\beta}, \mathbf{w}, \sigma^2}{\arg\max} \left\{ \begin{array}{l} -\frac{n}{2}\log(\sigma^2) - \frac{1}{2}\log(|\mathbf{R}|) \\ -\frac{1}{2\sigma^2}(\mathbf{y} - \mathbf{F}\boldsymbol{\beta})^T \mathbf{R}^{-1} (\mathbf{y} - \mathbf{F}\boldsymbol{\beta}) \end{array} \right\} \tag{29}$$

where $\log(\cdot)$ is the natural logarithm, $|\cdot|$ denotes the determinant operator, $\mathbf{y} = [y_{(1)}, \ldots, y_{(n)}]^T$ are the $n$ outputs in the training data, $\mathbf{R}$ is the correlation matrix with entries $R_{ij} = r(\mathbf{u}_{(i)}, \mathbf{u}_{(j)})$, and $\mathbf{F}$ is the prior mean basis matrix with entries $F_{ij} = f_j(\mathbf{u}_{(i)})$.

Once the parameters are estimated, the predicted response at the query point $\mathbf{u}^*$ is obtained via:

$$\hat{y}(\mathbf{u}^*) = f(\mathbf{u}^*)\hat{\boldsymbol{\beta}} + \mathbf{g}^T(\mathbf{u}^*)\mathbf{V}^{-1}(\mathbf{y} - \mathbf{F}\hat{\boldsymbol{\beta}}) \tag{30}$$

where $\mathbf{g}(\mathbf{u}^*)$ is an $n \times 1$ vector with the $i^{th}$ element $g_i(\mathbf{u}^*) = \hat{\sigma}^2 r(\mathbf{u}_{(i)}, \mathbf{u}^*)$, and $\mathbf{V}$ is the covariance matrix with entries $V_{ij} = \hat{\sigma}^2 r(\mathbf{u}_{(i)}, \mathbf{u}_{(j)})$.

### 4.5. Calibrations via LMGP

The detailed steps of the proposed framework are included in Algorithm 1. Our LMGP-based data-driven calibration has two major steps which are detailed below and demonstrated in see Section 5.4.

In the first step, we build the training dataset where the responses (UTS and toughness) characterize RVEs' effective softening behavior while the inputs are pore morphology descriptors, damage parameters, and simulation fidelity level. While the latter input is qualitative/categorical and is chosen based on the simulator cost, the other inputs are all quantitative and selected via DoE. For training sample $i$, we generate the RVE corresponding to the $i^{th}$ set of descriptors via descriptor-based reconstruction techniques and. We then deform this RVE via the simulator with the chosen fidelity level which uses the damage parameters of training sample $i$. Once the training dataset is built, we train the LMGP that simultaneously surrogates all the data sources.

In the second step, we solve an optimization problem to estimate the damage parameters that must be used for an ROM such that it predicts the same UTS and toughness as DNS which uses known material properties (i.e., a fixed set of values for $\bar{\mathrm{E}}^{cr}$ and $\alpha$) for any RVE. The estimated $\bar{\mathrm{E}}^{cr}$ and $\alpha$ for an ROM depend on the microstructural descriptors of the RVE (numerical inputs) and the ROM's fidelity level (a categorical input). Hence, the objective function of the (inverse) optimization problem measures the difference between predictions of ROM and DNS conditioned on these mixed inputs. Once $\bar{\mathrm{E}}^{cr}$ and $\alpha$ (i.e., the modified material properties) are estimated for each RVE for all ROMs, we conduct multiscale damage analyses where microscale simulations are carried out via ROMs.

---

**Algorithm 1** Framework of the data-driven calibration for ROM damage parameters via LMGP

```
1:  procedure Calibrate the damage parameters of ROMs with different morphologies and fidelity levels
2:      ▷ DoE variables include pore descriptors, damage model parameters, and simulation fidelity levels
3:      ▷ Fidelity level = 1, 2, and 3: use ROMs as simulators with different numbers of clusters (k).
4:      ▷ Fidelity level = 4: use DNS as the damage simulator
5:      ▷ Step-1:
6:      Set up upper and lower bounds of DoE variables and load DoE
7:      for i ← 1, N do    ▷ Loop over a total of N DoE samples
8:          Read pore descriptors at the DoE point-i
9:          ▷ MCR: microstructure characterization and reconstruction
10:         Reconstruct RVE's geometry via MCR based on pore descriptor values
11:         Load mesh module to generate FE mesh on the reconstructed RVE geometry
12:         ▷ CM: condensation method
13:         Load CM (Section 4.2) to compute the effective elastic modulus ℂ_M^el
14:         ▷ Assume the effective RVE properties are isotropic
15:         Compute Lame constants from ℂ_M^el
16:         Save the effective Lame constants of the RVE-i
17:         Read the damage parameters and fidelity level at the DoE sample point-i
18:         ▷ Damage responses include ultimate tensile strength (UTS) and toughness
19:         ▷ Perform damage analyses (Section 4.1)
20:         if fidelity level ← {1, 2 or 3}, then
21:             ▷ Use ROM for damage analyses
23:             Read k and load ROM (k) to compute effective damage responses (Section 4.3)
24:         elseif fidelity level ← 4, then
25:             ▷ Use DNS for damage analyses
26:             Load DNS to compute effective damage responses
27:         end
28:         Save the effective damage responses for each DoE sample point-i
29:      end for
30:      ▷ Step-2:
31:      Read effective damage responses
32:      Encode damage responses to let LMGP surrogate two damage responses (UTS and toughness)
33:      Load LMGP (Section 4.4)    ▷ Consider model fidelity levels as categorical variables
34:      ▷ Calibrate ROM damage parameters (Ē^cr and α) for different RVE and fidelity level
35:      for i ← 1, N do    ▷ Loop over N RVE samples
36:          for j ← {1, 2, 3} do    ▷ Loop over three different ROM fidelity levels
37:              ▷ Use damage parameters as optimization variables
38:              Minimize the difference of the damage responses between DNS_i and ROM_ij
39:              Save the optimal ROM damage parameters to a database
40:          end for
41:      end for
42:      return the database of the calibrated damage parameters of ROMs
```

43:    **end procedure**

## 5. Numerical studies

We apply the proposed data-driven framework to calibrate the ROMs in a multi-fidelity and multiscale model that simulates the damage behavior of a metallic component with spatially varying microstructures. In subsection 5.1, we train a GP to emulate the condensation method to accelerate the online calculation of $\mathbb{C}_M^{el}$ for each microstructure. In subsection 5.2, we construct a multi-fidelity model via LMGPs which are then used in subsection 5.3 to calibrate the damage parameters of ROMs. In subsection 5.4, we use the calibrated ROMs to investigate the influence of porosity on the structural damage responses of a multiscale model.

The material studied in this work is the cast aluminum alloy A356 whose elastic properties are:

$$Y = 5.70\text{e}4 \text{ MPa}, \quad v = 0.33 \quad (31)$$

where $Y$ and $v$ are Young's modulus and Poisson's ratio, respectively. The alloy's plasticity is modeled by following the J2 plasticity theory with the piecewise linear hardening curve in Figure 3. We assume that plasticity satisfies an associative plastic flow rule with the yield condition as:

$$\bar{S} \leq S_Y\left(\bar{E}^{pl}\right) \quad (32)$$

where $\bar{S}$, $\bar{E}^{pl}$ and $S_Y$ are Mises equivalent stress, equivalent plastic strain, and yield stress, respectively.

The softening behavior of A356 is modeled by the progressive damage model in Equation (10) with two damage parameters that are applied for all ROMs and DNS: critical plastic strain ($\bar{E}^{cr}$) and damage evolutionary rate parameter ($\alpha$). The two damage parameters are selected for calibration in this work because they both significantly affect damage responses; however, more parameters can be calibrated using our proposed framework.

$\bar{E}^{cr}$ determines the onset of softening that influences the largest stress that a material can withstand and $\alpha$ controls the amount of released fracture energy which determines the degradation rate of material properties amid damage evolution. The values of damage parameters used in DNS are given in Equation (33), while their values for ROMs need to be calibrated based on microstructural morphology and fidelity levels.

$$\bar{E}^{cr} = 0.03; \quad \alpha = 100 \quad (33)$$

Our method is implemented in Matlab [24] and we obtain the RVE responses on a high-performance cluster paralleled by 40 cores (AMD EPYC processor running at 4.1 GHz) with 120 GB RAM.

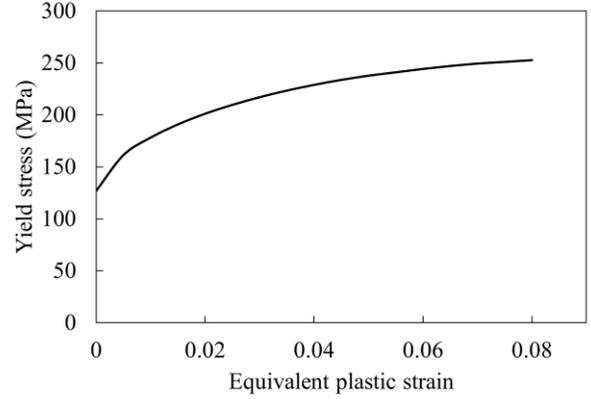

**Figure 3 Hardening behavior:** Piecewise linear hardening.

### 5.1. GP modeling for microstructure effective tangents

As described in Sections 4.1 and 4.2, the effective elastic tangent matrix relates the effective reference stresses with elastic strains. This matrix plays a fundamental role in continuum damage analysis since it enables simulating the progressive fracture evolutions at any IPs in a multiscale model, see Figure 1(a). However, computing the effective tangents often involves intensive computational efforts even when condensation methods are applied, see line 13 of Algorithm 1. Hence, we improve efficiency by developing a GP surrogate that correlates porosity morphology with microstructural effective tangent matrix.

We approximate the complex pores via overlapping ellipsoids whose geometry and spatial distribution are characterized by the following four descriptors: porosity volume fraction $V_f$, number of pores $N_p$, aspect ratio between ellipsoidal axes $A_r$, and the mean nearest distance between centroids $\bar{r}_d$. In addition, as we assume to work with isotropic microstructural responses, the components of the tangent matrix are reduced to two effective Lame constants ($\mu$ and $\lambda$). In this manner, our GP aims to build a predictive model between $[V_f, N_p, A_r, \bar{r}_d]$ and $[\mu, \lambda]$.

To construct the GP, we first generate a training dataset with 160 RVEs. The inputs in this dataset are generated via a DoE where each sample specifies the values of $[V_f, N_p, A_r, \bar{r}_d]$ for each RVE. We let the pore parameters satisfy the ranges in Equation (34) where L represents RVE's side length. Then, we use a microstructure reconstruction algorithm [25] to rebuild RVEs corresponding to DoE points. We demonstrate 12 reconstructed microstructures from the DoE datasets in Figure 4 where their pore descriptors $[V_f, N_p, A_r, \bar{r}_d]$ are enumerated in Table 1. In the reconstructed RVEs, the pore sizes are much smaller than the microstructures which reduces the property variations across different microstructure realizations that have the same four descriptors.

Once the dataset of RVEs is built, we use the condensation method to calculate the effective Lame constants for each RVE. Finally, we train a GP to emulate the relation between $[V_f, N_p, A_r, \bar{r}_d]$ and $[\mu, \lambda]$.

$$\begin{cases} 1\% \leq V_f \leq 20\% \\ 10 \leq N_p \leq 100 \\ 1 \leq A_r \leq 5 \\ 0.1L \leq \bar{r}_d \leq 0.5L \end{cases} \quad (34)$$

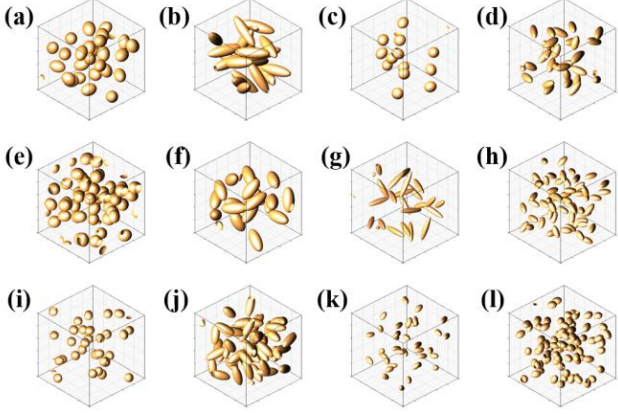

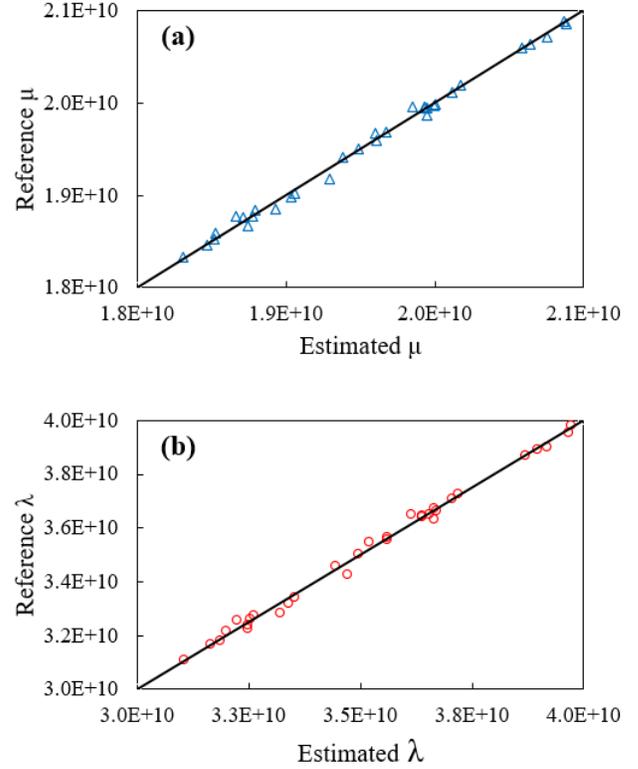

**Figure 4 Example reconstructed microstructures:** the values of microscale porosity descriptors and effective Lame constants are listed in Table 1.

**Table 1 Pore descriptors and effective Lame constants:** the numbers correspond to the reconstructed microstructures in Figure 4.

| RVE | $V_f$ | $N_p$ | $A_r$ | $\bar{r}_d$ | $\mu$ (e10) | $\lambda$ (e10) |
|---|---|---|---|---|---|---|
| (a) | 6.56% | 26 | 1.31 | 23.3 | 1.94 | 3.51 |
| (b) | 9.21% | 20 | 3.33 | 19.7 | 1.82 | 3.05 |
| (c) | 2.06% | 13 | 1.14 | 28.1 | 2.08 | 3.96 |
| (d) | 3.29% | 29 | 2.37 | 20.5 | 2.03 | 3.78 |
| (e) | 9.97% | 48 | 1.16 | 20.4 | 1.85 | 3.23 |
| (f) | 7.80% | 20 | 2.15 | 25.9 | 1.89 | 3.31 |
| (g) | 1.92% | 22 | 4.95 | 22.4 | 2.08 | 3.92 |
| (h) | 3.12% | 60 | 2.11 | 16.9 | 2.04 | 3.81 |
| (i) | 2.61% | 31 | 1.09 | 21.6 | 2.07 | 3.91 |
| (j) | 9.70% | 51 | 2.47 | 18.2 | 1.82 | 3.09 |
| (k) | 1.15% | 36 | 1.84 | 21.1 | 2.11 | 4.03 |
| (l) | 4.48% | 77 | 1.43 | 14.5 | 2.01 | 4.02 |

To test the GP's accuracy, we split the dataset by using 80% for training and 20% for validation. Comparisons of the predictions with the test samples are shown in Figure 5.

**Figure 5 Emulation accuracy:** comparison of actual microstructural effective Lame constants against GP predictions on unseen test samples.

To assess the convergence and whether sufficient training data are used, we split the dataset into 100 samples for training and 60 samples for testing. We sequentially increase the size of the training data from 10 to 100 and evaluate the accuracy of the corresponding GPs on 60 test samples (all GPs are evaluated on the same set of test samples). The prediction errors are computed by Equation (35):

$$E_{\mathbf{y}} = \frac{1}{N_v} \sum_{i=1}^{N_v} \frac{\|\hat{\mathbf{y}}_i - \bar{\mathbf{y}}_i\|}{\|\bar{\mathbf{y}}_i\|} \quad (35)$$

where $N_v$ is the number of validation samples, $E_{\mathbf{y}}$ is the relative prediction error of responses $\mathbf{y} = [\mu, \lambda]$, $\hat{\mathbf{y}}_i$ and $\bar{\mathbf{y}}_i$ are the predicted effective Lame constants of the $i^{th}$ RVE. The convergence curve is shown in Figure 6 where it is observed that with the increase of training samples, prediction errors monotonically decrease. With 100 samples the prediction error drops to lower than 0.4%, indicating highly accurate predictions. Therefore, we use the GP emulated effective modulus to replace the condensation method amid online computations to accelerate damage analyses for all microstructures.

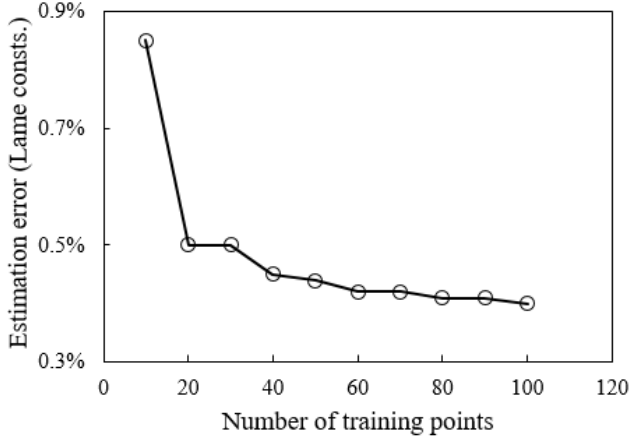

**Figure 6 Error convergence:** GP estimation errors on predicted Lame constants with respect to the number of training points.

## 5.2. LMGP modeling of damage parameters

In this subsection, we train an LMGP that is used in subsection 5.3 for ROM calibration, see Figure 1(c) where the LMGP is used in the inverse optimization. We first demonstrate the importance of calibration and then provide the details on the training and validation of the LMGP.

To demonstrate the importance of parameter calibrations of ROMs, let us consider the microstructure in Figure 7(a) with pore descriptors $[V_f, N_p, A_r, \bar{r}_d] = [15.9\%, 25, 1.4, 24.3]$ and damage parameters in Equation (33). We subject this RVE to the deformation gradient in Equation (36) and compute its responses via the DNS using 68675 finite elements as shown in Figure 7(b) where significant strain concentrations are observed in the vicinity of pores. We then model this RVE via an ROM with 3200 clusters using the same damage parameters as the DNS. The plastic strain distributions are shown in Figure 7(c) which demonstrates the diffusive nature of local clustering.

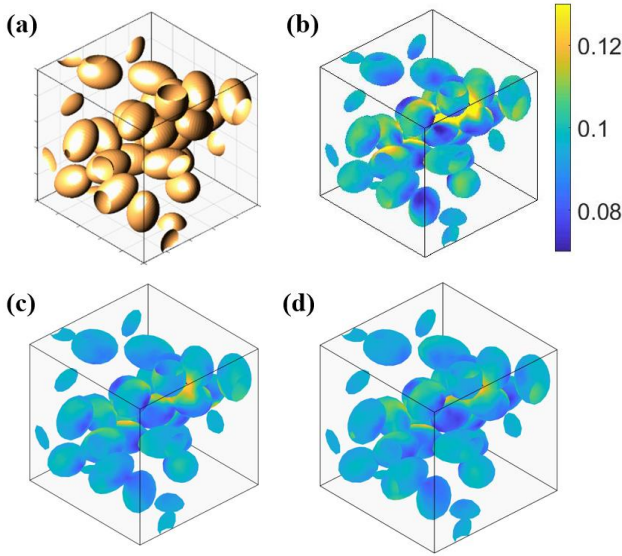

**Figure 7 Equivalent plastic strain fields: (a)** the porosity morphology of a microstructure with 25 pores, **(b)** plastic strains are simulated via DNS, **(c)** plastic strains are approximated by ROM (k=3200) without calibration, and **(d)** plastic strains are approximated by ROM (k=3200) with calibration.

$$\mathbf{F}^M = \begin{bmatrix} 1.1 & 0 & 0 \\ 0 & 0.95 & 0 \\ 0 & 0 & 0.95 \end{bmatrix} \quad (36)$$

Comparison of strain distributions in Figure 7 demonstrates that the label of the data source (i.e., DNS or ROM), which we consider as a categorical variable in LMGP, must encode the diffusive nature of the local solutions. Additionally, compared to the DNS, the cluster-wise solutions of ROMs have lower magnitudes of plastic strains which result in delayed fracture initiation, higher UTS, and larger material toughness, see Figure 8(a) and Table 2.

**Table 2 Damage responses:** values of the UTS and toughness of DNS and ROMs for the microstructure in Figure 7.

| Simulation fidelity | Pre-calibration | | After calibration | |
|---|---|---|---|---|
| | UTS | Toughness | UTS | Toughness |
| DNS | 1.03e8 | 3.71e6 | - | - |
| k=800 | 1.15e8 | 4.10e6 | 1.07e8 | 3.85e6 |
| k=1600 | 1.09e8 | 3.93e6 | 1.043e8 | 3.783e6 |
| k=3200 | 1.06e8 | 3.85e6 | 1.046e8 | 3.776e6 |

The accuracy of ROMs can be improved by calibrating their damage parameters ($\bar{E}^{cr}$ and $\alpha$). We illustrate calibration effects on the local strain concentrations and effective behaviors in Figure 7(d) and Figure 8(b), respectively. It is evident that compared to the ROMs with the damage parameters of DNS, the calibrated ROMs provide more accurate and effective stress-strain responses. Specifically, the calibration algorithm calibrates (i.e., reduces) the ROMs' critical plastic strain to induce early softening so that their UTS values become closer to that observed in DNS. Meanwhile, the calibration also decreases the ROMs' damage evolutionary rate parameters to compensate for the toughness reduction due to early softening.

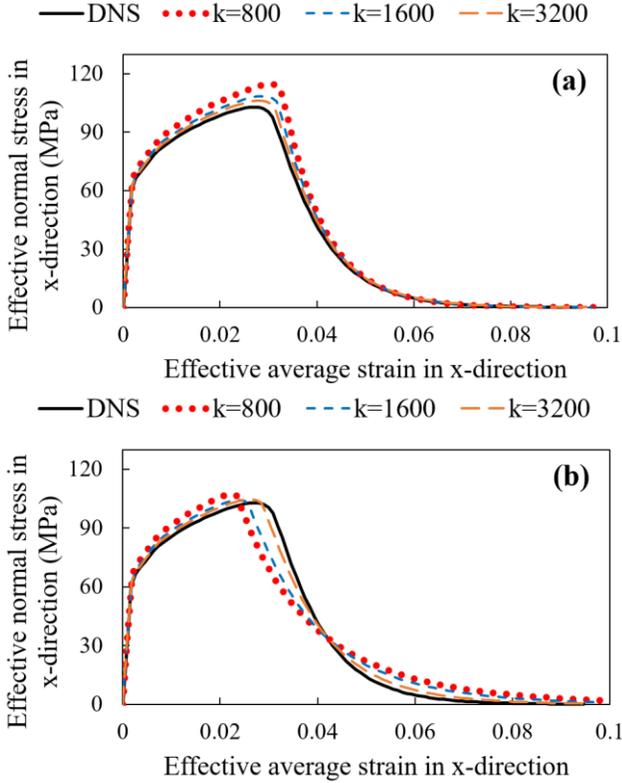

**Figure 8 Importance of calibration: (a)** effective stress-strain curves without damage parameters calibration, and **(b)** the effective response with calibration.

We further compare the values of the material toughness and UTS between DNS and ROMs in Table 2 where we find that the accuracies of both damage responses are improved after calibrations. The improvement is demonstrated by the enumerated errors in Table 3 where we observe that the calibrations significantly reduce the ROMs' model errors ($r$) for all ROM fidelity levels ($k = 800, 1600, 3200$). Additionally, the magnitudes of the normalized errors ($\|r\| = \|r_{toughness}, r_{UTS}\|_2$) continuously drop with the increase of clusters, which validates our observation in Figure 8 that the ROM with more clusters provides closer solutions to the DNS in both pre and post-calibration scenarios.

**Table 3 ROM prediction error:** errors of ROMs on UTS and toughness for the microstructure in Figure 7.

| ROM clusters ($k$) | Errors ($r$) w/o LMGP calibration (%) | | | Errors ($r$) w. LMGP calibration (%) | | |
|---|---|---|---|---|---|---|
| | UTS | Toughness | $\|r\|$ | UTS | Toughness | $\|r\|$ |
| 800 | 11.5 | 10.6 | 15.61 | 3.74 | 3.64 | 5.22 |
| 1600 | 5.5 | 5.8 | 7.99 | 1.28 | 1.96 | 2.34 |
| 3200 | 3.2 | 3.7 | 4.86 | 1.55 | 1.73 | 2.33 |

We provide the values of the calibrated damage parameters in Table 4. We note that with the decrease of simulation fidelity levels from DNS to the ROMs ($k = 3200, 1600, 800$), both values of $\overline{E}^{cr}$ and $\alpha$ decrease. This trend implicitly validates our previous observation in Figure 7(b)-(c) that fewer clusters result in more diffusive cluster-wise plastic strains with delayed damage initiations in Figure 8(a). This is because, to counteract the artificial delay of softening, the calibration algorithm needs to lower $\overline{E}^{cr}$ so that softening occurs at smaller deformation conditions. Meanwhile, the calibration decreases $\alpha$ to lower the material's degradation rates during damage evolutions which helps to approximate ROMs' toughness to that of DNS.

**Table 4 Calibrated damage parameters:** values of calibrated ROM damage parameters for the microstructure in Figure 7.

| Simulation fidelity | $\overline{E}^{cr}$ | $\alpha$ |
|---|---|---|
| ROM ($k$=800) | 0.021 | 36.31 |
| ROM ($k$=1600) | 0.024 | 47.65 |
| ROM ($k$=3200) | 0.027 | 72.27 |
| DNS | 0.03 | 100 |

Once the accuracy of ROMs is improved via calibration, they can substitute DNS. We compare the computational time of DNS against the ROMs in Figure 9 by CPU time. We observe that while DNS takes 29.8 hours to finish the damage simulation, it only takes about 68.8, 27.9 and 15.6 minutes for the ROMs with 3200, 1600, and 800 clusters, respectively. The ROMs' computational costs can be further reduced to online time since their offline stages (clustering and preprocessing) are performed only once and are unnecessary in future calculations. Thus, by comparing the online time with DNS, the acceleration factors of the ROMs with 3200, 1600, and 800 clusters are 44.1, 99.9, and 242.9, respectively.

After demonstrating the necessity of ROM calibration, we now describe the proposed LMGP-based calibration approach. Compared to manual calibrations, the proposed data-driven approach is highly efficient in automatically allocating the optimal values of the damage parameters for the ROMs based on their fidelity levels as well as the microstructure.

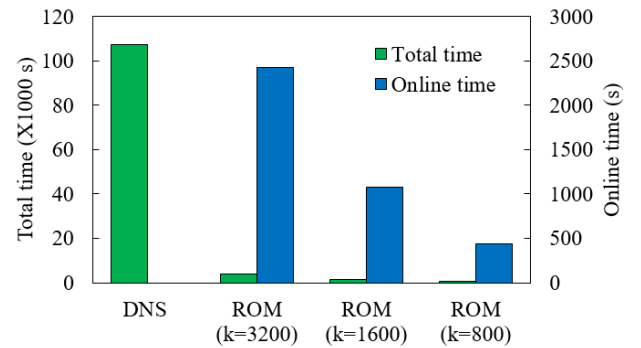

**Figure 9 Time Reduction:** computational time comparison between DNS with ROMs.

To use LMGP for calibration, we generate a dataset consisting of six inputs $\boldsymbol{x} = [x_1, \ldots, x_6]^T$ and two outputs $\boldsymbol{y}$, as shown in Table 5. The first four inputs represent the pore descriptors (i.e., $[V_f, N_p, A_r, \bar{r}_d]$) and the last two inputs represent the two damage parameters (i.e., evolutionary rate parameter $\alpha$ and critical effective plastic strain $\bar{E}^{cr}$). We generate DoE sample points via Sobol sequence by satisfying the ranges of descriptor values and damage parameters in Equation (34) and (37), respectively. Two LMGP outputs are the two damage responses (UTS and material toughness).

$$\begin{cases} 1\% \leq \bar{E}^{cr} \leq 3\% \\ 10 \leq \alpha \leq 100 \end{cases} \quad (37)$$

We append each sample point with a categorical variable to encode the data source which is denoted by $t_1 = \{1,2,3,4\}$ where label 4 corresponds to DNS while labels 3, 2, and 1 correspond to the ROM with $k = 3200$, $k = 1600$, and $k = 800$ respectively. To enable LMGP to simultaneously surrogate two damage responses, we also appended the samples with a second categorical variable encoding the type of responses by $t_2 = \{1, 2\}$ where label 1 corresponds to UTS and label 2 indicates material toughness. Part of the resulting single-response training dataset is shown in Table 5. Our entire dataset contains 600 samples that are created from four different fidelity sources: 70, 110, 170, and 250 samples are from DNS and the ROMs with 3200, 1600, and 800 clusters respectively.

**Table 5 Training dataset of LMGP:** four microstructure descriptors ($x_1 \sim x_4$), two damage parameters ($x_5 \sim x_6$), and two categorical inputs ($t_1 \sim t_2$) which encode data source and response type. The data are color-coded based on $t_2$ (green is UTS and blue toughness).

| $x_1$ | $x_2$ | $x_3$ | $x_4$ | $x_5$ | $x_6$ | $t_1$ | $t_2$ | $y$ |
|---|---|---|---|---|---|---|---|---|
| 0.021 | 13 | 1.14 | 28.1 | 54.7 | 0.015 | 4 | 1 | 1.12e8 |
| ⋮ | ⋮ | ⋮ | ⋮ | ⋮ | ⋮ | ⋮ | ⋮ | ⋮ |
| 0.066 | 26 | 1.31 | 23.3 | 71.2 | 0.017 | 4 | 1 | 1.15e8 |
| 0.098 | 87 | 1.89 | 12.4 | 75.6 | 0.020 | 3 | 1 | 1.13e8 |
| ⋮ | ⋮ | ⋮ | ⋮ | ⋮ | ⋮ | ⋮ | ⋮ | ⋮ |
| 0.045 | 77 | 1.43 | 14.5 | 80.7 | 0.023 | 3 | 1 | 1.26e8 |
| 0.030 | 70 | 3.93 | 12.6 | 73.4 | 0.066 | 2 | 1 | 1.21e8 |
| ⋮ | ⋮ | ⋮ | ⋮ | ⋮ | ⋮ | ⋮ | ⋮ | ⋮ |
| 0.026 | 31 | 1.10 | 21.6 | 98.3 | 0.029 | 2 | 1 | 1.33e8 |
| 0.078 | 34 | 2.77 | 17.4 | 21.3 | 0.012 | 1 | 1 | 1.08e8 |
| ⋮ | ⋮ | ⋮ | ⋮ | ⋮ | ⋮ | ⋮ | ⋮ | ⋮ |
| 0.016 | 88 | 3.13 | 14.4 | 61.7 | 0.027 | 1 | 1 | 1.36e8 |
| 0.021 | 13 | 1.14 | 28.1 | 54.7 | 0.015 | 4 | 2 | 3.14e6 |
| ⋮ | ⋮ | ⋮ | ⋮ | ⋮ | ⋮ | ⋮ | ⋮ | ⋮ |
| 0.067 | 26 | 1.31 | 23.3 | 71.2 | 0.017 | 4 | 2 | 3.00e6 |
| 0.098 | 87 | 1.89 | 12.4 | 75.6 | 0.020 | 3 | 2 | 3.26e6 |
| ⋮ | ⋮ | ⋮ | ⋮ | ⋮ | ⋮ | ⋮ | ⋮ | ⋮ |
| 0.045 | 77 | 1.43 | 14.5 | 80.7 | 0.023 | 3 | 2 | 3.93e6 |
| 0.030 | 70 | 3.93 | 12.6 | 73.4 | 0.066 | 2 | 2 | 3.07e6 |
| ⋮ | ⋮ | ⋮ | ⋮ | ⋮ | ⋮ | ⋮ | ⋮ | ⋮ |
| 0.026 | 31 | 1.10 | 21.6 | 98.3 | 0.029 | 2 | 2 | 4.72e6 |
| 0.078 | 34 | 2.77 | 17.4 | 21.3 | 0.012 | 1 | 2 | 3.17e6 |
| ⋮ | ⋮ | ⋮ | ⋮ | ⋮ | ⋮ | ⋮ | ⋮ | ⋮ |
| 0.016 | 88 | 3.13 | 14.4 | 61.7 | 0.027 | 1 | 2 | 5.05e6 |

To investigate the effects of sample sizes on prediction accuracy, we fit our LMGP to 100, 200, 300, and 400 samples and test its performance on 200 testing samples across 50 random repetitions. We note that our dataset is unbalanced because we have fewer samples from high-fidelity source that requires high computational costs and much more data points from low-fidelity models. In particular, 10% of the training samples are obtained from DNS, 20% from ROM with 3200 clusters, 30% from the ROM with 1600 clusters, and 40% from ROM with 800 clusters. These ratios are the same across all sub-datasets that are created using the entire dataset.

We scale LMGP outputs to $[0, 1]$ and compute the mean absolute errors (MAE) of predictions as shown in Figure 10. We observe that with the increase of training samples, both MAE and its variance decrease. Therefore, we choose 400 samples as our training dataset which contains 40 samples of DNS, 80, 120, and 160 samples from the ROMs with 3200, 1600, and 800 clusters, respectively. Based on users' computational budgets, various combinations of different fidelity sources can be explored. Minimizing the costs of training datasets for multi-fidelity models is, however, out of the scope of this work.

From Figure 10, we also notice that the scale of the vertical axis in Figure 10(b) is smaller than that of Figure 10(a), suggesting that our LMGP provides better predictions for toughness than UTS (we elaborate on the underlying reasons below).

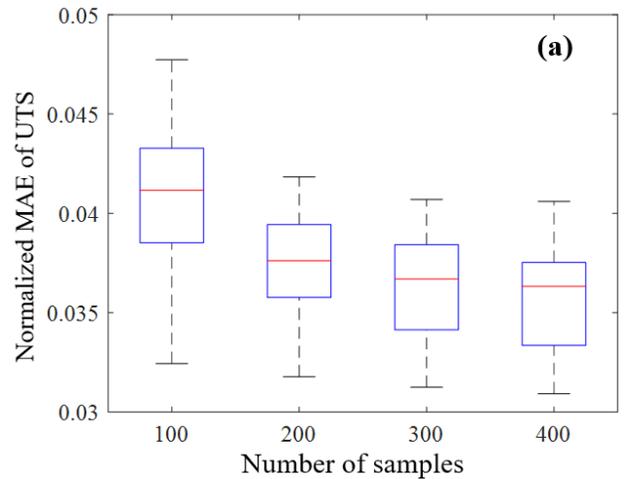

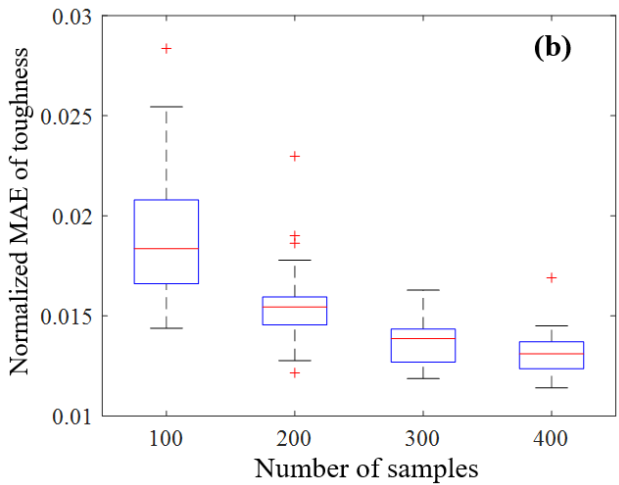

**Figure 10 LMGP's MAEs:** normalized MAE of UTS and toughness with respect to different numbers of training samples.

Once LMGP is trained, we can visualize the learned latent space where each combination of the two categorical variables is mapped to a latent position in Figure 11. Based on the latent points of the underlying [fidelity level, response] combinations, it is evident that latent axes $z_1$ and $z_2$ encode, respectively, the types of damage response and the simulation fidelity levels (note that this encoding is learned automatically by LMGP). We observe that the scale of $z_1$ is one order of magnitude larger than $z_2$, suggesting that the latent points are primarily grouped by their damage responses ($z_1$). For the same damage response, the latent points are further distinguished by their fidelity levels ($z_2$). Specifically, we find that the positions of $k = 3200$ are far from $k = 800$ but close to DNS, suggesting the damage responses predicted by ROMs with 3200 clusters share more similarity with the DNS than the ROMs with only 800 clusters. This also indicates that low fidelity models (e.g., $k = 800$) exhibit model form error compared to the higher fidelity sources.

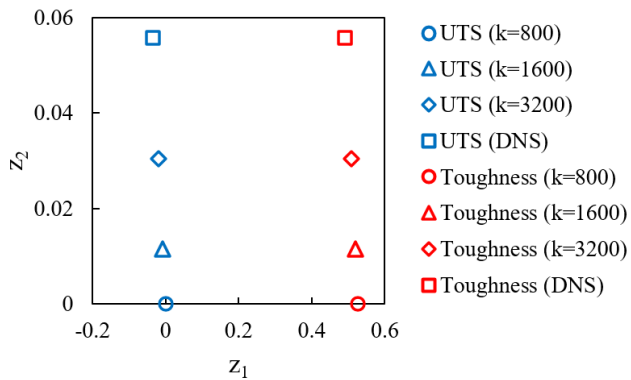

**Figure 11 Learnt latent space of LMGP:** each latent position encodes simulation fidelity level and damage response.

LMGP provides significant insights and interpretations of the characteristics of the studied datasets in Figure 11. For example, the vertical distance between DNS and $k = 800$ along $z_2$ direction is about $0.06$; suggesting a correlation value of ($\exp\{-0.06^2\} = 0.9964$) between the two data sources, see Equation (27). Given this correlation, LMGP can use the knowledge from the low-fidelity data (i.e., $k = 800$) to improve its accuracy in emulating the high-fidelity source (i.e., DNS). We also notice that the horizontal distance along $z_1$ axis is about $0.6$, resulting in the correlation value of $\exp\{-0.6^2\} = 0.6977$. It suggests the two responses are positively correlated, consistent with our expectation that the delayed fracture initiation from ROMs' diffusive local solutions not only increases predicted UTS but also enlarges material toughness, see the discussion in Figure 8(a).

To assess LMGP's accuracy, we split the 600 sample points into training and validation sets where 400 samples are used for training and the remaining 200 samples are for validation. The validation dataset contains 20, 40, 60, and 80 samples from DNS, the ROM with $k = 3200$, 1600, and 800, respectively. LMGP's prediction accuracy is quantified by the MAE in Table 6 where it is observed that the prediction errors are higher for the highest fidelity source (DNS) as well as the lowest-fidelity ROM with 800 clusters compared to the other two fidelity sources. The reason for the large prediction errors on DNS comes from its data scarcity, and the errors of $k = 800$ is due to its inherent model errors.

**Table 6 Error Analysis:** MAE of the LMGP's prediction for the two damage responses and four data sources.

| Simulation fidelity | MAE | |
|---|---|---|
| | UTS | Toughness |
| ROM with k = 800 | 0.0430 | 0.0152 |
| ROM with k = 1600 | 0.0277 | 0.0104 |
| ROM with k = 3200 | 0.0274 | 0.0100 |
| DNS | 0.0444 | 0.0283 |

We plot LMGP's predictions against validation samples for the two damage responses in Figure 12 where the predictions of both responses are found to be quite accurate. Specifically, we notice that the predictions of UTS have higher errors than those of toughness. This observation is consistent with our discussions in Figure 10. One plausible reason is that UTS, as a point measurement of the maximum stress that an RVE can tolerate, is sensitive to some important factors that are not considered in this surrogate, e.g., the directions of crack propagations. In contrast, RVE toughness, which is a global estimation for the amount of released fracture energy amid damage evolution (which is an integral quantity), is characterized by our model quite well.

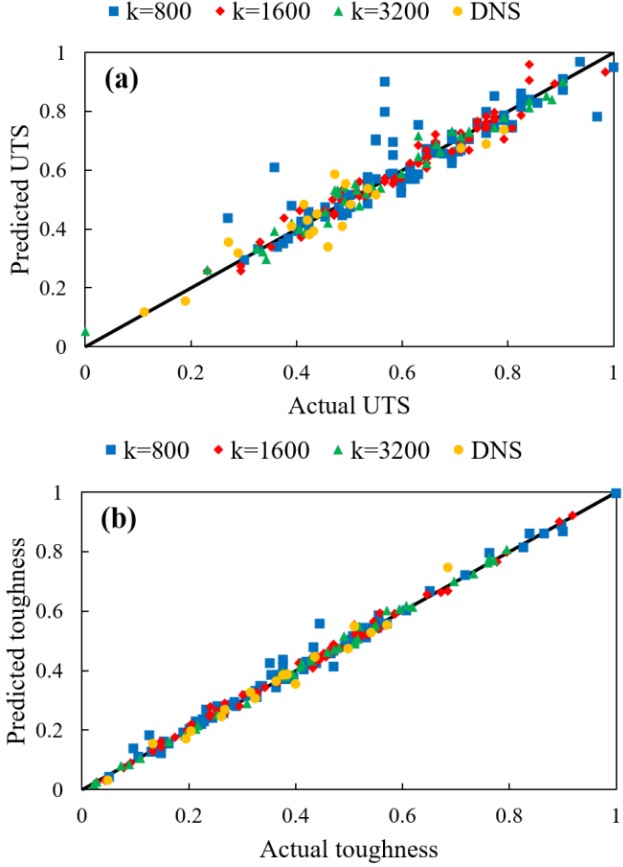

**Figure 12 Performance on unseen test data:** comparison of the true responses against the LMGP's predictions for UTS and toughness.

### 5.3. Calibration of damage parameters

To improve solution accuracy, the damage parameters of ROMs need to be calibrated as shown in Figure 1(c). We perform the calibration by solving an inverse optimization problem whose objective function is evaluated via LMGP. We estimate the calibration parameters for the $i^{th}$ microstructure and the $j^{th}$ source-level such that the estimated damage responses from ROM match the ones from DNS that uses the true damage parameters: $\alpha_{DNS} = 100$ and $\bar{E}_{DNS}^{cr} = 0.03$. The optimization problem is hence formulated as:

$$\left[\hat{\alpha}, \bar{E}^{cr}\right] = \arg\min_{\alpha, \bar{E}^{cr}} \left\| y_p\left(x_{DNS}^i\right) - y_p\left(x_j^i\right) \right\|^2 \quad (38)$$

where $y_p(\cdot)$ are the predicted damage responses by LMGP and $x_{DNS}^i = [V_f^i, N_p^i, A_r^i, \bar{r}_d^i, \alpha_{DNS}, \bar{E}_{DNS}^{cr}, t_1 = 4, t_2]$ represents the input vector of the $i^{th}$ microstructure for predicting the responses of DNS. $x_j^i = [V_f^i, N_p^i, A_r^i, \bar{r}_d^i, \alpha, \bar{E}^{cr}, t_1 = j, t_2]$ is the input vector of the $i^{th}$ microstructure for predicting the damage responses for ROM at the $j^{th}$ fidelity level (note that we pass $t_2$ as a vector to get both damage responses).

We use a gradient-based optimization method to solve Equation (38). In Figure 13, we demonstrate the values of the calibrated damage parameters for the 600 microstructures in the database. We note that the calibration is performed based on an inverse optimization which tends to minimize the difference between the damage responses of DNS and ROM where both are surrogated by LMGP, that is, no on-line microstructure simulation is performed for calibration. Since the optimization relies on an inexpensive surrogate, its computational cost is very small.

From Figure 13, we observe the same trend across all samples. Specifically, we highlight the calibrated damage values of two distinct RVEs that were not used in training the LMGP. We find that: (i) the values of the ROM's calibrated damage parameters are smaller than those of DNS (represented by dashed lines), and (ii) the values of calibrated damage parameters are closer to the DNS' values as we increase the number of clusters ($k$). For instance, the calibrated parameters of both RVEs with 800 clusters are much smaller than their counterparts in DNS or ROMs with 1600 or 3200 clusters. The underlying reason is that as $k$ decreases, the ROM's local plastic strain becomes more diffusive which delays damage initiation and artificially increases UTS and toughness. Therefore, to counteract this diffusive behavior, the calibrated damage parameters tend to reduce the strength of the materials to induce early damage such that the ROMs can faithfully approximate DNS.

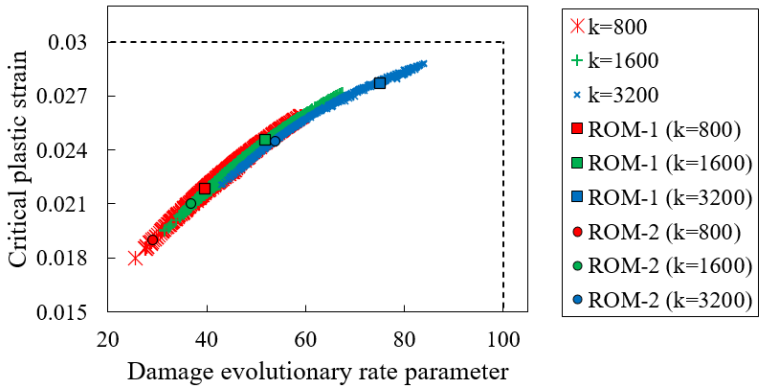
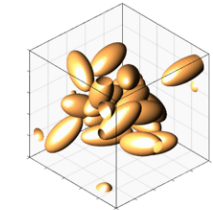
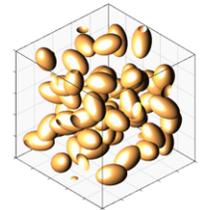

ROM-1: [13.7%, 21, 2.13, 22.3]   ROM-2: [17.3%, 57, 1.58, 17.7]

**Figure 13 Calibrated damage parameters:** calibrated damage parameters of 600 samples simulated by ROMs with three fidelity levels where two RVEs with distinct pore morphologies are highlighted.

## 5.4. Multiscale damage analyses

Since manufacturing-induced porosity significantly affects material properties [26–28], in this section we can apply the reduced multiscale damage model to a 3D L-shape bracket to quantify the impact of micro-porosity on the bracket's fracture behavior. Our simulations follow Figure 1(d) where the calibrated ROMs are used to accelerate the microscale analyses in the multiscale model.

The dimensions of the L-bracket are shown in Figure 14. The bracket is fixed on the top surface, and it is subject to a Dirichlet boundary condition on the right surface ($d = 20$mm). The bracket model is discretized with 2113 linear tetrahedron elements with reduced integrations.

For multiscale analysis, we divide the bracket into two subdomains: a monoscale region and a multiscale region with spatially varying porosity distribution. This choice is motivated by the observation that under large deformations the fracture happens in the multiscale domain where high accuracy and microstructural effects are needed, and hence the other regions of the bracket can be modeled as a single scale.

For each of the 147 IPs in the multiscale region, we randomly assign a microstructure from the database generated in Section 5.2. The effective damage behavior in each microstructure is simulated by ROMs with three fidelity options: 800, 1600, or 3200 clusters. For each ROM with a selected cluster number, its optimal damage parameters are readily available from the LMGP-based calibration process described in Section 5.3. Specifically, among the 147 macro-IPs, 77 IPs are associated with the RVEs simulated by 800 clusters, 50 and 20 IPs are assigned to the RVEs with 1600 and 3200 clusters, respectively. We note that the RVEs with higher numbers of clusters are assigned to the IPs with anticipated softening that is predicted by a preliminary single scale simulation without micro-pores, see the two highlighted RVEs with distinct local pore morphologies that are assigned to different IPs in the multiscale region in Figure 14.

In our multiscale simulations, we ensure the released fracture energy is consistent between the scales by equating microstructure volumes to macroscopic mesh sizes. Additionally, we apply a nonlocal damage function with a feature size of 15mm on the bracket model to prevent pathological mesh dependency and convergence difficulty.

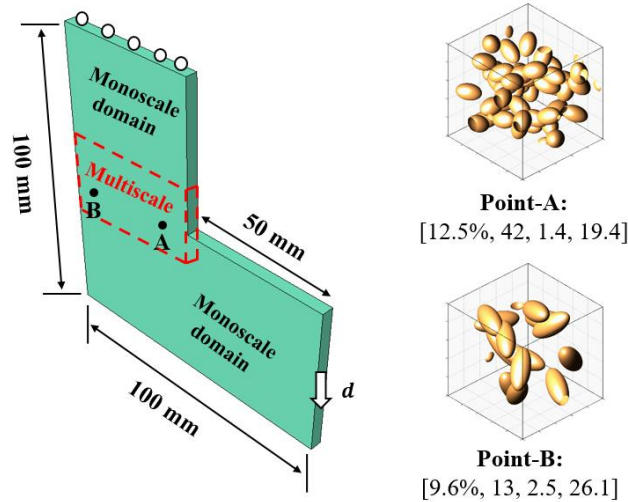

**Figure 14 Multiscale model:** the dimensions and boundary conditions of a 3D L-shape bracket model with a thickness of 5 mm where two RVEs with distinct pore descriptors are associated with two macroscale IPs in the multiscale domain.

We demonstrate the simulated fracture pattern, local plastic strain distributions, and load-displacement response (with and without multiscale treatment) in Figure 15. In Figure 15(a), we demonstrate the macro-fractures by elements' effective damage values $D_M$ in Equation (12) where $D_M = 1$ represents complete material ruptures. We notice that the highlighted two macro-IPs are located in the damage zone, and we plot the distributions of microscopic equivalent plastic strains in Figure 15(b) exhibiting significant local strain concentrations. Specifically, we observe large plastic strains are accumulated in proximity to pore surfaces in the two RVEs which cause the macroscale fractures in Figure 15(a). In Figure 15(c), we observe that porous microstructures significantly deteriorate the bracket's load-carrying capacity which drops by 10.22% from 70.86N to 63.62N, and the bracket breaks at a much lower displacement boundary condition. Therefore, compared to the single-scale model that only considers dense materials and neglects pores, the multiscale model provides us with a more realistic prediction by considering fractures across scales.

Our multiscale simulation is paralleled by 40 CPU cores on a high-performance HPC, and it is finished in 15.2 hours. Based on the efficiency comparison between the ROM and DNS in Figure 9, the estimated computational time for DNS (classic FE$^2$) is more than 2623.5 hours (109.3 days), that is, our calibrated ROM speeds up the overall computation compared to DNS with an acceleration factor of 172.6.

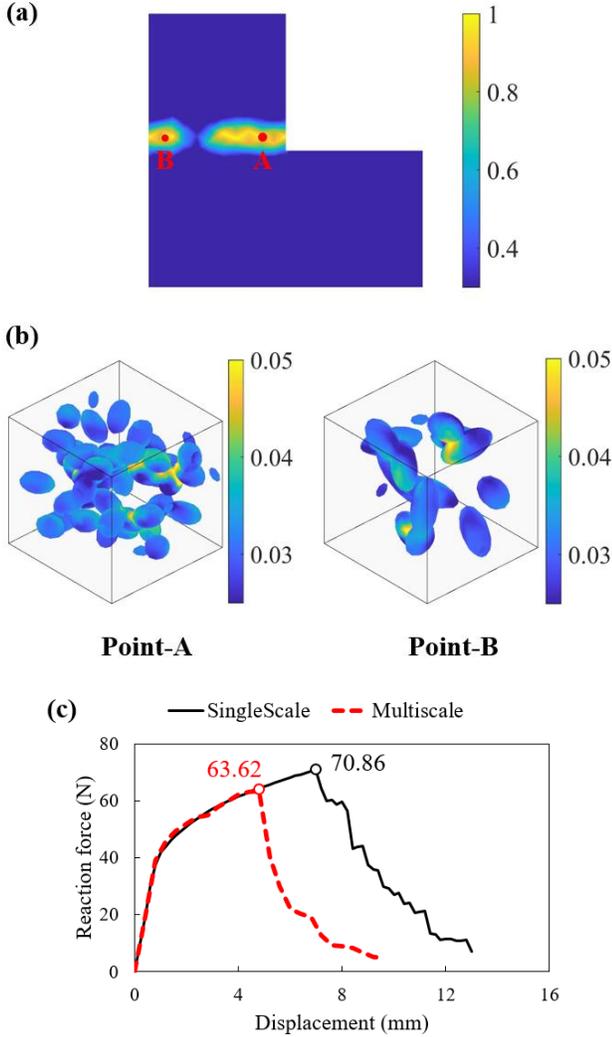

**Figure 15** Results of multiscale damage analysis: **(a)** the top view of the fracture patterns on the L-bracket model, **(b)** the distributions of equivalent plastic strains in the two highlighted RVEs, and **(c)** the force-displacement responses.

## 6. Conclusion

We propose a multi-fidelity reduced-order model for multiscale damage analysis that considers manufacturing-induced spatially varying porosity. Our model is not only significantly faster than multiscale simulations based on the $FE^2$ approach, but also has lower memory requirements. Our approach relies on a mechanics-based reduced-order model (ROM) that accelerates the microscale elasto-plastic deformations by clustering the degrees of freedom. Since this clustering artificially increases microstructures' tolerance to damage initiation and evolution, we develop a calibration scheme to estimate the damage parameters that must be used in ROM such that it can faithfully approximate high-fidelity simulations.

We employ latent map Gaussian processes (LMGPs) to build a multi-fidelity emulator which is then used in our calibration scheme. In addition to providing high accuracy and versatility for emulation, we show that the learned latent space of LMGP is interpretable and provides insights into the problem (e.g., determining the relative accuracy of multiple ROMs with respect to DNS). This LMGP-based calibration scheme differs from existing calibration works such as [29, 30] which focus on calibrating simulations using experimental data. In contrast to these works, we focus on calibrating ROMs against DNS such that these ROMs can be used in multiscale simulations where microstructural details vary over the macro-component.

In this work, we use the calibrated ROMs in a multiscale simulation to study the effect of spatially varying micro-porosity on the macroscopic response of an L-bracket model. Our results indicate that porosity noticeably decreases the strength of the material and hence must be considered in "design for fracture".

In this work, we neglect the inherent uncertainty in material properties, i.e., our simulations (based on either ROMs or DNS) are deterministic. More realistic fracture modeling requires embedding uncertainty sources in our calibration scheme and multiscale simulations. With this treatment, we will obtain probabilistic distributions for the calibrated parameters (conditioned on a selected fidelity level). These microscale distributions are spatially correlated at the macroscale and quantifying their effects on the macroscale quantities relies on sampling technique [31] (e.g., based on Markov chain Monte Carlo). We believe our ROMs provide a unique opportunity for such sampling-based multiscale uncertainty quantification and plan to pursue this direction in our future works.

The proposed data-driven multi-fidelity damage model in this paper opens up some interesting future research directions. For example, the ROMs with calibrated damage parameters can efficiently generate material response databases correlating intricate microstructural morphologies with effective material behaviors under complex loading conditions. Such databases enable deep learning-based surrogates for a direct mapping between material local morphology and their responses for computationally demanding nonlinear analyses. In addition, applying our multi-fidelity model to investigate the effects of material uncertainty on structural behaviors is vital for robust designs as engineered material systems are inherently embedded with manufacturing-induced uncertainties that propagate across scales.

### Acknowledgments

We acknowledge the support from National Science Foundation (award number OAC-2103708) as well as Early Career Faculty grant from NASA's Space Technology Research Grants Program (award number 80NSSC21K1809). The authors also thank the ACRC consortium members for their support.

**References**

[1] N. Oune and R. Bostanabad, "Latent map Gaussian processes for mixed variable metamodeling," *Computer Methods in Applied Mechanics and Engineering*, vol. 387, p. 114128, Dec. 2021, doi: 10.1016/j.cma.2021.114128.

[2] J. T. Eweis-Labolle, N. Oune, and R. Bostanabad, "Data Fusion With Latent Map Gaussian Processes," *Journal of Mechanical Design*, vol. 144, no. 9, Jun. 2022, doi: 10.1115/1.4054520.

[3] G. J. Dvorak, "Transformation field analysis of inelastic composite materials," *Proceedings of the Royal Society of London. Series A: Mathematical and Physical Sciences*, vol. 437, no. 1900, pp. 311–327, May 1992, doi: 10.1098/rspa.1992.0063.

[4] J. C. Michel and P. Suquet, "Nonuniform transformation field analysis," *International Journal of Solids and Structures*, vol. 40, no. 25, pp. 6937–6955, Dec. 2003, doi: 10.1016/S0020-7683(03)00346-9.

[5] S. Roussette, J. C. Michel, and P. Suquet, "Nonuniform transformation field analysis of elastic–viscoplastic composites," *Composites Science and Technology*, vol. 69, no. 1, pp. 22–27, Jan. 2009, doi: 10.1016/j.compscitech.2007.10.032.

[6] Z. Liu, M. A. Bessa, and W. K. Liu, "Self-consistent clustering analysis: An efficient multi-scale scheme for inelastic heterogeneous materials," *Computer Methods in Applied Mechanics and Engineering*, vol. 306, pp. 319–341, Jul. 2016, doi: 10.1016/j.cma.2016.04.004.

[7] G. Cheng, X. Li, Y. Nie, and H. Li, "FEM-Cluster based reduction method for efficient numerical prediction of effective properties of heterogeneous material in nonlinear range," *Computer Methods in Applied Mechanics and Engineering*, vol. 348, pp. 157–184, May 2019, doi: 10.1016/j.cma.2019.01.019.

[8] S. Deng, C. Soderhjelm, D. Apelian, and R. Bostanabad, "Reduced-order multiscale modeling of plastic deformations in 3D alloys with spatially varying porosity by deflated clustering analysis," *Comput Mech*, Jun. 2022, doi: 10.1007/s00466-022-02177-8.

[9] Z. Liu, M. Fleming, and W. K. Liu, "Microstructural material database for self-consistent clustering analysis of elastoplastic strain softening materials," *Computer Methods in Applied Mechanics and Engineering*, vol. 330, pp. 547–577, Mar. 2018, doi: 10.1016/j.cma.2017.11.005.

[10] C. Miehe, "Numerical computation of algorithmic (consistent) tangent moduli in large-strain computational inelasticity," *Computer Methods in Applied Mechanics and Engineering*, vol. 134, no. 3, pp. 223–240, Aug. 1996, doi: 10.1016/0045-7825(96)01019-5.

[11] V. Kouznetsova, W. A. M. Brekelmans, and F. P. T. Baaijens, "An approach to micro-macro modeling of heterogeneous materials," *Computational Mechanics*, vol. 27, no. 1, pp. 37–48, Jan. 2001, doi: 10.1007/s004660000212.

[12] S. Deng, C. Soderhjelm, D. Apelian, and R. Bostanabad, "Reduced-Order Multiscale Modeling of Plastic Deformations in 3D Alloys with Spatially Varying Porosity by Deflated Clustering Analysis," *Computational Mechanics, accepted*, doi: 10.1007/s00466-022-02177-8.

[13] S. Tang, L. Zhang, and W. K. Liu, "From virtual clustering analysis to self-consistent clustering analysis: a mathematical study," *Comput Mech*, vol. 62, no. 6, pp. 1443–1460, Dec. 2018, doi: 10.1007/s00466-018-1573-x.

[14] M. R. Ackermann, J. Blömer, D. Kuntze, and C. Sohler, "Analysis of Agglomerative Clustering," *Algorithmica*, vol. 69, no. 1, pp. 184–215, May 2014, doi: 10.1007/s00453-012-9717-4.

[15] A. Likas, N. Vlassis, and J. J. Verbeek, "The global k-means clustering algorithm," *Pattern Recognition*, vol. 36, no. 2, pp. 451–461, Feb. 2003, doi: 10.1016/S0031-3203(02)00060-2.

[16] G. R. Liu, *Mesh Free Methods: Moving Beyond the Finite Element Method*, 1st edition. CRC Press, 2002.

[17] T. Jönsthövel, M. B. Gijzen, C. Vuik, C. Kasbergen, and A. Skarpas, "Preconditioned Conjugate Gradient Method Enhanced by Deflation of Rigid Body Modes Applied to Composite Materials," *CMES - Computer Modeling in Engineering and Sciences*, vol. 47, pp. 97–118, Jul. 2009.

[18] R. Aubry, F. Mut, S. Dey, and R. Löhner, "Deflated preconditioned conjugate gradient solvers for linear elasticity," *International Journal for Numerical Methods in Engineering*, vol. 88, no. 11, pp. 1112–1127, 2011, doi: 10.1002/nme.3209.

[19] P. Yadav and K. Suresh, "Large Scale Finite Element Analysis Via Assembly-Free Deflated Conjugate Gradient," *Journal of Computing and Information Science in Engineering*, vol. 14, no. 4, Oct. 2014, doi: 10.1115/1.4028591.

[20] C. E. Rasmussen, "Gaussian Processes in Machine Learning," in *Advanced Lectures on Machine Learning: ML Summer Schools 2003, Canberra, Australia, February 2 - 14, 2003, Tübingen, Germany, August 4 - 16, 2003, Revised Lectures*, O. Bousquet, U. von Luxburg, and G. Rätsch, Eds. Berlin, Heidelberg: Springer, 2004, pp. 63–71. doi: 10.1007/978-3-540-28650-9_4.

[21] R. Bostanabad, Y.-C. Chan, L. Wang, P. Zhu, and W. Chen, "Globally Approximate Gaussian Processes for Big Data With Application to Data-Driven Metamaterials Design," *Journal of Mechanical Design*, vol. 141, no. 11, Sep. 2019, doi: 10.1115/1.4044257.

[22] R. Planas, N. Oune, and R. Bostanabad, "Extrapolation With Gaussian Random Processes and Evolutionary Programming," presented at the ASME 2020 International Design Engineering Technical Conferences and Computers and Information in Engineering Conference, Nov. 2020. doi: 10.1115/DETC2020-22381.

[23] R. Planas, N. Oune, and R. Bostanabad, "Evolutionary Gaussian Processes," *Journal of Mechanical Design*, vol. 143, no. 11, May 2021, doi: 10.1115/1.4050746.

[24] "'MATLAB. (2010). version 7.10.0 (R2010a). Natick, Massachusetts: The MathWorks Inc.'"

[25] R. Bostanabad *et al.*, "Computational microstructure characterization and reconstruction: Review of the state-of-the-art techniques," *Progress in Materials Science*, vol. 95, pp. 1–41, Jun. 2018, doi: 10.1016/j.pmatsci.2018.01.005.

[26] S. Deng, C. Soderhjelm, D. Apelian, and K. Suresh, "Estimation of elastic behaviors of metal components containing process induced porosity," *Computers & Structures*, vol. 254, p. 106558, Oct. 2021, doi: 10.1016/j.compstruc.2021.106558.

[27] S. Deng, C. Soderhjelm, D. Apelian, and K. Suresh, "Second-order defeaturing estimator of manufacturing-induced porosity on structural elasticity," *International Journal for Numerical Methods in Engineering*, vol. n/a, no. n/a, doi: 10.1002/nme.7042.

[28] S. Deng, D. Apelian, and R. Bostanabad, "Concurrent Multiscale Damage Analysis with Adaptive Spatiotemporal Dimension Reduction." arXiv, May 24, 2022. doi: 10.48550/arXiv.2205.12149.

[29] B. D. Youn, B. C. Jung, Z. Xi, S. B. Kim, and W. R. Lee, "A hierarchical framework for statistical model calibration in engineering product development," *Computer Methods in Applied Mechanics and Engineering*, vol. 200, no. 13, pp. 1421–1431, Mar. 2011, doi: 10.1016/j.cma.2010.12.012.

[30] A. Olleak and Z. Xi, "Calibration and Validation Framework for Selective Laser Melting Process Based on Multi-Fidelity Models and Limited Experiment Data," *Journal of Mechanical Design*, vol. 142, no. 8, Feb. 2020, doi: 10.1115/1.4045744.

[31] R. Bostanabad *et al.*, "Uncertainty quantification in multiscale simulation of woven fiber composites," *Computer Methods in Applied Mechanics and Engineering*, vol. 338, pp. 506–532, Aug. 2018, doi: 10.1016/j.cma.2018.04.024.